\documentclass{ifacconf}  

\usepackage{natbib,graphicx}
\usepackage{amsmath}
\usepackage[hang,raggedright]{subfigure}
\usepackage{psfrag}
\usepackage{amssymb}
\usepackage{latexsym}
\usepackage{amsmath}
\usepackage{subfigure}
\usepackage{algorithm}
\usepackage{algorithmic}
\usepackage{color}

\newtheorem{remark}[thm]{\bf Remark}
\newtheorem{definition}{\bf Definition}[section]

\def\qed{\hfill\rule[-1pt]{5pt}{5pt}\par\medskip}

\begin{document}

\begin{frontmatter}

\title{Probabilistically Safe Vehicle Control in a Hostile Environment\thanksref{footnoteinfo}} 

\thanks[footnoteinfo]{This work was supported in part by grants ONR-MURI N00014-09-1051, ARO W911NF-09-1-0088, AFOSR YIP FA9550-09-1-020, NSF CNS-0834260, and the United Technologies Research Center.}

\author[First]{Igor Cizelj} 
\author[Second]{Xu Chu (Dennis) Ding} 
\author[Second]{Morteza Lahijanian}
\author[Third]{Alessandro Pinto}
\author[Second]{Calin Belta}

\address[First]{Division of Systems Engineering, Boston University, Boston, MA 02215, USA. (e-mail: icizelj@bu.edu)}
\address[Second]{Department of Mechanical Engineering, Boston University, Boston, MA 02215, USA. (e-mail: {\{xcding; morteza; cbelta\}@bu.edu)}}                                              
\address[Third]{Embedded Systems and Networks Group, United Technologies Research Center Inc., Berkeley, CA (e-mail: alessandro.pinto@utrc.utc.com)}

\begin{abstract}                    
In this paper we present an approach to control a vehicle in a hostile environment with static obstacles and moving adversaries.  The vehicle is required to satisfy a mission objective expressed as a temporal logic specification over a set of properties satisfied at regions of a partitioned environment.  We model the movements of adversaries in between regions of the environment as Poisson processes.  Furthermore, we assume that the time it takes for the vehicle to traverse in between two facets of each region is exponentially distributed, and we obtain the rate of this exponential distribution from a simulator of the environment. We capture the motion of the vehicle and the vehicle updates of adversaries distributions as a Markov Decision Process.   Using tools in Probabilistic Computational Tree Logic, we find a control strategy for the vehicle that maximizes the probability of accomplishing the mission objective.  We demonstrate our approach with illustrative case studies. 
\end{abstract}

\end{frontmatter}

\section{Introduction}
\label{sec:intro}
Robot motion planning and control has been widely studied in the last twenty years.  Recently, temporal logics, such as Linear Temporal Logic (LTL) and Computational Tree Logic (CTL) have become increasingly popular for specifying robotic tasks (see, for example, \citep{cooner:valet,karaman:vehicle,kloetzer:fully,loizou:automatic}).  It has been shown that temporal logics can serve as rich languages capable of specifying complex mission tasks such as ``go to region A and avoid region B unless regions C or D are visited''.

Many of the above-mentioned works that use a temporal logic as a specification language rely on the assumption that the motion of the robot in the environment can be abstracted to a finite transition system by partitioning the environment.   The transition system must be finite in order to allow the use of existing model-checking tools for temporal logics (see \citep{baier:principles}). Furthermore, it is assumed that the resultant transition system obtained from the abstraction process is deterministic ({\it i.e.}, an available control action deterministically triggers a unique transition from one region of the environment to anther region), and the environment is static.  To address environments with dynamic obstacles, \citep{kress-gazit:whereswaldo?,topcu:receding} find control strategies that guarantee satisfactions of specifications by playing temporal logic games with the environment.

In practice, due to noise introduced in control (actuator error) or the environment (measurement error), a deterministic transition system may not adequately represent the motion of the robot. \citep{kloetzer:dealing} proposed a control strategy for a purely non-deterministic transition system ({\it i.e.}, a control action enables multiple possible transitions to several regions of the environment).   \citep{lahijanian:motion} pushed this approach a step further by modeling the motion of the robot as a Markov Decision Process (MDP) ({\it i.e.}, a control action triggers a transition from one region to anther with some fixed and known probability).  The transition probabilities of this MDP can be obtained from empirical  measurements or an accurate simulator of the environment. A control strategy was then derived to satisfy a mission task specified in Probabilistic Computational Tree Logic (PCTL) with the maximum probability.

In this paper, we extend this approach to control a vehicle in a dynamic and threat-rich environment with static obstacles and moving adversaries.   We assume that the environment is partitioned into polygonal regions, and a high level mission objective is given over some properties assigned to these regions.  We model the movements of adversaries in a region as Poisson processes. Furthermore, we model the time it takes for the vehicle to reach from one facet of a region to another facet as an exponential random variable.  This motion model is supported by our realistic simulator of the environment, and we obtain the rate of this exponential random variable from the simulator.

The main contribution of this paper is an approach to design a reactive control strategy that provides probabilistic guarantees of accomplishing the mission in a threat-rich environment. This control strategy is reactive in the sense that the control of the vehicle is updated whenever the vehicle reaches a new region in the environment, or an adversary moves in between the current region and its adjacent region ({\it i.e.} if the vehicle observes movements of adversaries, it updates the adversary distributions for adjacent regions and chooses a different control action as needed).  In order to solve this problem, we capture the motion of the vehicle, as well as vehicle estimates of the adversary distributions in a MDP. This way, we map the vehicle control problem to the problem of finding a control policy for an MDP such that the probability of satisfying a PCTL formula is maximized. For latter, we use our previous approach presented in \citep{lahijanian:motion}.

The method that we propose here is closely related to ``classical" Dynamic Programming (DP) - based approaches \citep{alterovitz:stochastic}. In particular, it can be seen as a simple extension of a Maximum Reachability Probability (MRP) problem, which itself is a simple case of a stochastic shortest path (SSP) problem \citep{bertsekas:dynamic}. In these problems, the set of allowed specifications is restricted to reaching a given destination state, and the corresponding optimal control strategy is found by solving one linear program (LP). In contrast, our proposed PCTL control framework allows for richer, temporal logic specifications and multiple destinations. In addition, through the use of nested probabilities, it allows for specifying sub-task probabilities.

The rest of the paper is organized as follows.  Sec. \ref{sec:prelim} introduces necessary notations, definitions and preliminary results.  Sec. \ref{sec:problem} formulates the problem and describes our approach.  Sec. \ref{sec:MDP} describes the construction process of the MDP modelling the vehicle in the environment.  Sec. \ref{sec:vehicle_control} explains how we generate the desired vehicle control strategy.  A simulator of the vehicle environment is detailed in Sec. \ref{sec:simulator} and some numerical case studies are shown in Sec. \ref{sec:examples}.   Sec. \ref{sec:conclusions} concludes the paper.

\section{Preliminaries}
\label{sec:prelim}
\subsection{Markov Decision Process and Probability Measure}
\begin{definition}[Markov Decision Process (MDP)]
\label{def:MDP}
A labeled MDP $\mathcal{M}$ is a tuple $(S,s_0,Act,A, P,\Pi,h)$ where 
\begin{itemize}
\item $S$ is a finite set of states;
\item $s_0 \in S$ is the initial state; 
\item $Act$ is a set of actions;
\item $A: S\rightarrow 2^{Act}$ is a function specifying the enabled actions at a state $s$;
\item $P: S \times Act \times S \rightarrow [0,1]$ is a transition probability function such that for all states $s \in S$ and actions $a \in A(s)$: $\sum_{s' \in S}P (s,a,s')=1$, and for all actions $a\notin A(s)$ and $s'\in S$, $P (s,a,s')=0$;
\item $\Pi$ is the set of properties;
\item  $h: S \rightarrow 2^{\Pi}$ is a function that assigns some properties in $\Pi$ to each state of $s\in S$.
\end{itemize}
\end{definition}

A path $\omega$ through an MDP is a sequence of states $\omega=s_0s_1\ldots s_is_{i+1}\ldots$ where each transition is induced by a choice of action at the current step $i$. We denote the set of all finite paths by $\text{Path}^{fin}$ and of infinite paths by $\text{Path}$. 

\begin{definition}[Control Policy]
\label{def:ConPol}
A control policy $\mathcal{G}$ of an MDP model $\mathcal{M}$ is a function mapping a finite path $\omega^{fin} = s_0s_1s_2\hdots s_n$ of $\mathcal{M}$ onto an action in $A(s_n)$. In other words, a policy is a function $\mathcal{G} : \text{Path}^{fin} \rightarrow Act$ that specifies for every finite path, the next action to be applied. 
\end{definition}

Under policy $\mathcal{G}$, an MDP becomes an infinite discrete-time Markov Chain, denoted by $\mathcal{M}_{\mathcal{G}}$. Let path $\text{Path}_{\mathcal{G}} \subseteq \text{Path}$ and $\text{Path}_\mathcal{G}^{fin} \subseteq \text{Path}^{fin}$ denote the set of infinite and finite paths that can be produced under $\mathcal{G}$. Because there is a one-to-one mapping between $\text{Path}_\mathcal{G}$ and the set of paths of $\mathcal{M}_{\mathcal{G}}$ the Markov Chain induces a probability measure over $\text{Path}_\mathcal{G}$ as follows.  First, define a measure $\text{Pr}_{\mathcal{G}}^{fin}$ over the set of finite paths by setting the probability of $\omega^{fin} \in \text{Path}_{\mathcal{G}}^{fin}$ equal to the product of the corresponding transition probabilities in $\mathcal{M}_{\mathcal{G}}$.
Then, define $C(\omega^{fin})$ as the set of all (infinite) paths $\omega \in \text{Path}_{\mathcal{G}}$ with the prefix $\omega^{fin}$. The probability measure on the smallest $\sigma$-algebra over $\text{Path}_{\mathcal{G}}$ containing $C(\omega^{fin})$ for all $\omega^{fin} \subseteq \text{Path}_{\mathcal{G}}^{fin}$ is the unique measure satisfying 
\begin{equation}
\label{eq:measure}
\text{Pr}_{\mathcal{G}}(C(\omega^{fin}))=\text{Pr}_{\mathcal{G}}^{fin}(\omega^{fin}),
\end{equation}
for all $\omega^{fin} \in \text{Path}_{\mathcal{G}}^{fin}$.
\begin{figure}
            \centering
	\includegraphics[scale=0.45]{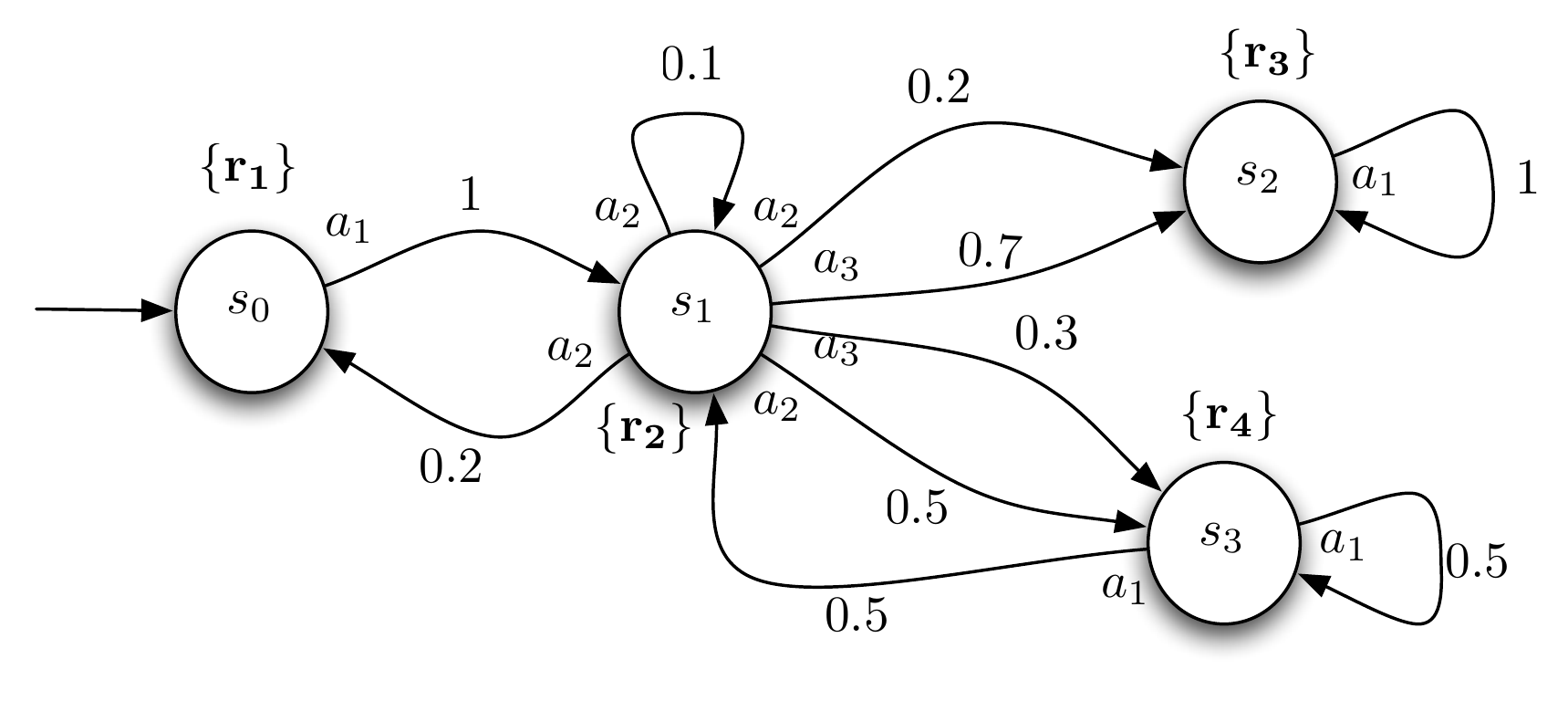}
      \caption{ 
      Example of a four-state MDP.  $A(s)$ and $h(s)$ are shown for each state.  The labels over the transitions correspond to the transition probabilities. Assume the simple control policy $\mathcal G$ defined by mapping: $\mathcal{G}(s_0)=a_1$, $\mathcal{G}(\cdots s_1)=a_2$, $\mathcal{G}(\cdots s_2)=a_1$ and $\mathcal{G}(\cdots s_3)=a_1$ where $\cdots s_i$ denotes any finite path terminating in $s_i$. It is easy to see that the probability of a finite path $s_0s_1s_1$ is $\text{Pr}_{\mathcal{G}}^{fin}(s_0s_1s_1)=0.1$. Under $\mathcal{G}$, the cylinder set of all infinite paths with this prefix is $C(s_0s_1s_1)=\{\overline{s_0s_1s_1}, s_0s_1s_1\overline{s_2},s_0s_1s_1\overline{s_3},s_0\overline{s_1},\ldots\}$. According to Eq. (\ref{eq:measure}), we have that $\text{Pr}_{\mathcal{G}}(C(s_0s_1s_1))=\text{Pr}_{\mathcal{G}}^{fin}(s_0s_1s_1)=0.1.$}
            \label{fig:Markov}
   \end{figure}
A simple MDP is shown in Fig. \ref{fig:Markov} to illustrate the above concepts.  We refer readers to \citep{baier:principles,ross:introduction} for more information about MDPs and probability measures defined on paths of an MDP.

\subsection{Probabilistic Computational Tree Logic}
Probabilistic Computational Tree Logic (PCTL) (\cite{rutten:mathematical}) is a probabilistic extension of CTL that includes the probabilistic operator $\mathcal{P}$.  PCTL formulas are interpreted either as truth values (true or false) or qualitative expressions (\emph{i.e.} find the maximum probability) of properties of the MDP.  Formulas are constructed by connecting properties from a set of properties $\Pi$ using Boolean operators ($\neg$ (negation), $\wedge$ (conjunction), and $\rightarrow$ (implication)), temporal operators ($\bigcirc$ (next), $\mathcal{U}$ (until)), and the probabilistic operator $\mathcal{P}$. This allows to express rich specifications given in natural language as PCTL formulas.

For example, consider the MDP shown in Fig. \ref{fig:Markov} and specification $\phi=\mathcal{P}_{max=?}[\neg r_3 \:  \mathcal{U}  \:  r_4 ]$. In words, this formula asks for the maximum probability of reaching the state satisfying $r_4$ ({\it i.e.,} $s_{3}$) without passing through the state satisfying $r_3$ ({\it i.e.,} $s_{2}$). This problem can be translated to a problem of finding the maximum probability of reaching a set of states of the MDP, using the probability measure of paths under a policy defined in the previous sub-section (for more details, see \cite{baier:principles,lahijanian:motion}). There are  probabilistic model-checking tools, such as PRISM (see \citep{kwiatkowska:probabilistic}), that solve this problem. More complex specifications can be obtained by nesting the probability operator and temporal operators, {\it e.g.,} the formula $\mathcal{P}_{max=?} [\neg r_3 \, \mathcal{U} \, (r_4 \wedge \mathcal{P}_{\geq 0.5}[\neg r_3 \, \mathcal{U} \, r_1])$], asks for the maximum probability of eventually visit state $s_3$ and then with probability greater than $0.5$ state $s_0$ while always avoiding state $s_2$.
\section{Problem formulation and approach}
\label{sec:problem}

\begin{figure*}[thpb]
\begin{center}
	\subfigure[A realistic scenario representing a city environment partitioned into regions.  $r_{p}$ denotes the ``pick-up'' region, and $r_{d}$ denotes the ``drop-off'' region.]
	{
		\label{fig:real_scenario}	
		\includegraphics[scale=.3]{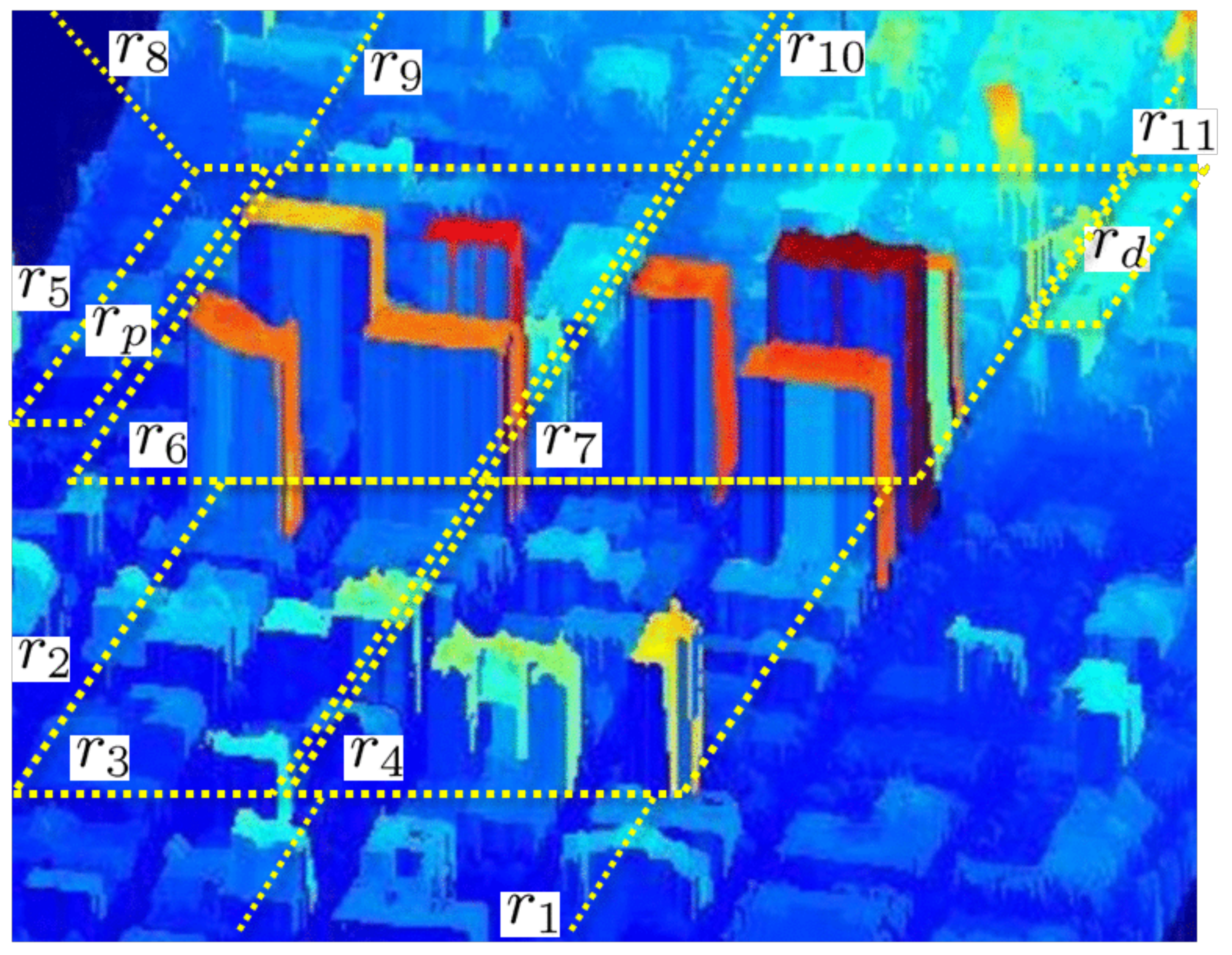}
	}	
	\subfigure[Possible motion of the vehicle in the environment.  The arrows represent movements of the vehicle in between facets, $\it{e.g.}$, the vehicle can choose to go from $f_{2}$ towards $f_5$.  For this scenario, we assume that, only at the pick-up and drop-off regions, the vehicle can enter and leave through the same facet.]{
		\label{fig:partitioned_environment}
		\includegraphics[scale=0.4]{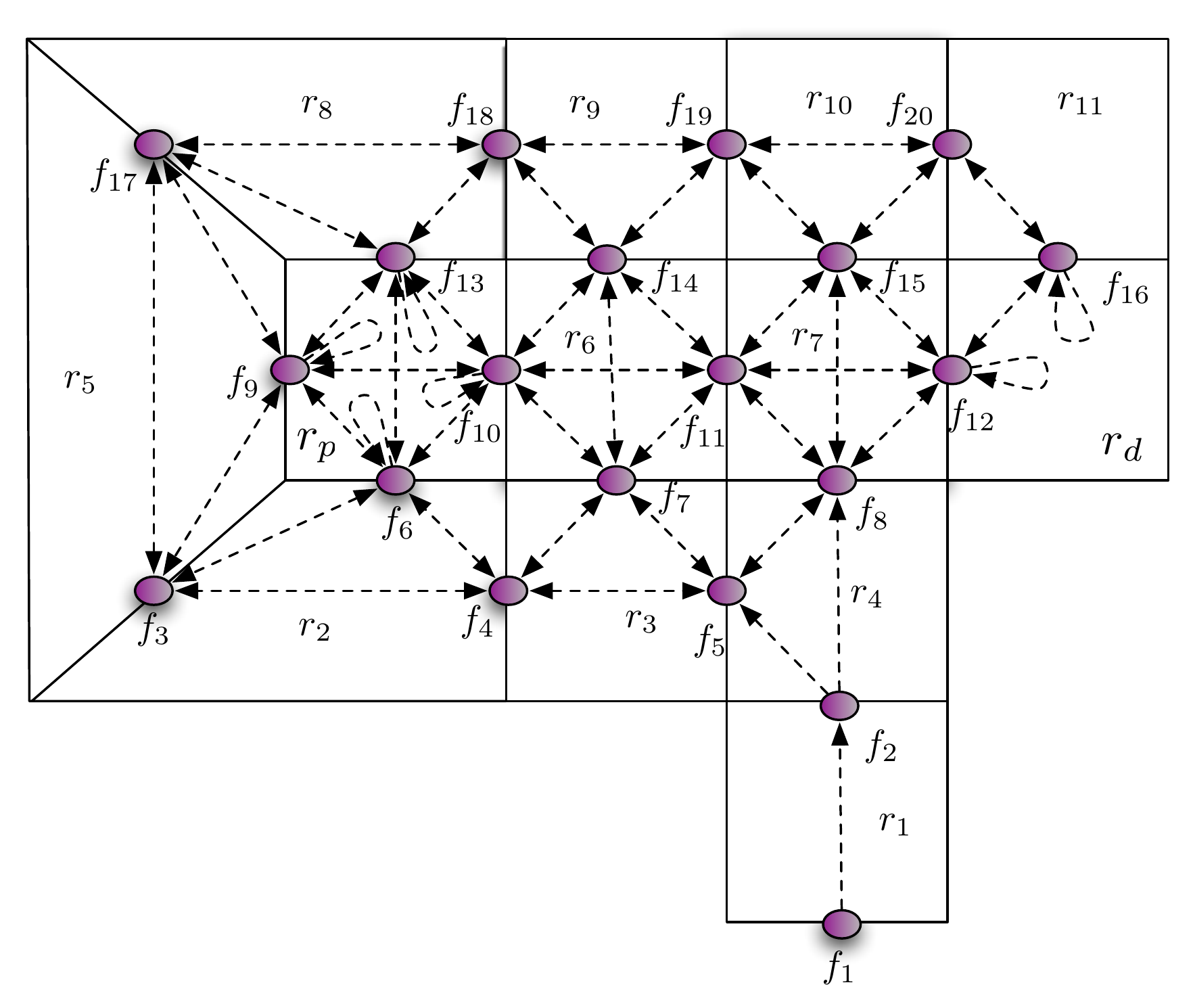}		
	}	
	\caption{Example of a partitioned city environment}
\end{center}
\end{figure*}

We consider a city environment that is partitioned into a set of  polytopic regions $R$.  We assume the partition\footnote{Throughout the paper, we relax the notion of a partition by allowing regions to share facets}is such that adjacent regions in the
environment share exactly one facet.  We denote $F$ as the set of facets of all polytopes in $R$.   We assume that one region $r_{p}\in R$ is labeled as the ``pick-up'' region, and another region $r_{d}\in R$ is labeled as the ``drop-off'' region.  Fig. \ref{fig:real_scenario} shows an example of a partitioned city environment.  We assume that there is a vehicle moving in the environment.  We require this vehicle to carry out the following mission objective: 

{\bf Mission Objective}: Starting from an initial facet $f_{init}\in F$ in a region $r_{init}\in R$, the vehicle is required to reach the pick-up region $r_{p}$ to pick up a load.  Then, the vehicle is required to reach the drop-off region $r_{d}$ to drop-off the load.

We consider a threat-rich environment with dynamic adversaries and static obstacles in some regions.  The probability of safely crossing a region depends on the number of adversaries and the obstacles in that region.  We say that the vehicle is lost in a region if it fails to safely cross the region (and thus fails the mission objective).  We assume that there is no adversary or obstacle in the initial region. 

Let integers $M_{r}$ and $N_{r}$ be the minimum and maximum number of adversaries in region $r\in R$, respectively.  We define 
\begin{equation}
\label{eq:probinit}
p^{init}_r: \{M_{r},\ldots,N_{r}\} \rightarrow [0,1]
\end{equation}
 as a given (initial) probability mass function for adversaries in region $r \in R$, {\it i.e.} $p^{init}_r(n)$ is the probability of having $n$ adversaries in region $r$ and $\sum_{n=M_{r}}^{N_{r}}p^{init}_{r}(n)=1$. However, adversaries may move in between regions.  We model the movements of adversary in a region by arrivals of customers in a queue.  Thus we consider the movements of adversary as Poisson processes and we assume that the time it takes for an adversary to leave and enter region $r$ is exponentially distributed with rate $\mu_{l}(r)$ and $\mu_{e}(r)$, respectively.  We further assume that adversaries move independent of each other, and at region $r$, the distributions of adversaries in adjacent regions of $r$ depend only on the adversaries in $r$ and the movements of adversaries between $r$ and its adjacent regions.
 
In addition, each region has an attribute that characterizes the presence of obstacles, which we call obstacle density.  We define 
\begin{equation}
p_r^o: \{0,1,\ldots,N^{o}_{r}\} \rightarrow [0,1],
\end{equation}
as the probability mass function of the obstacle density in region $r \in R$, $\it{i.e.}$, $p_r^o(o)$ is the probability of having obstacle density $o$ in region $r$ and $\sum_{o=0}^{N^{o}_{r}}(o)=1$.  Unlike adversaries, we assume that obstacles can not move in between regions.

We assume that the vehicle has a map of the environment and can detect its current region.  When the vehicle enters a region, it observes the number of adversaries and the obstacle density in this region.  When the vehicle is traversing inside a region, it detects movements of adversaries between the current region and its adjacent regions.  

The motion capability of the vehicle in the environment is limited by a (not necessarily symmetric) relation $\Delta \subseteq F \times F$, with the following meaning: If the vehicle is at a facet $f\in F$ and $(f,f') \in \Delta$, then it can use a motion primitive to move from $f$ towards $f'$ (without passing through any other facet), {\it i.e.,} $\Delta$ represents a set of motion primitives for the vehicle.  The control of the vehicle is represented by $(f,f')\in \Delta$, with the meaning that at facet $f$, $f'$ is the next facet the vehicle should move towards.  Fig. \ref{fig:partitioned_environment} shows possible motions of the vehicle in this environment.
We assume that the time it takes for the vehicle to move from facet $f$ to facet $f'$ is exponentially distributed with rate $\lambda(\delta)$, where $\delta=(f,f')\in \Delta$.   This assumption is based on results from a simulator of the environment (see Sec. \ref{sec:simulator}). 

During the time when the vehicle is executing a mission primitive $(f,f')$ ({\it i.e.,} moving between facet $f$ and $f'$), we denote the probability of losing the vehicle as:
\begin{equation}
\label{eq:probkill}
p_{\delta}^{lost}: \{M_{r},\ldots,N_{r}\} \times \{0,\ldots,N^{o}_{r}\} \rightarrow [0,1],
\end{equation}
where $\delta=(f,f') \in \Delta$, and $r$ is the region bounded by $f$ and $f'$.  We obtain $p_{\delta}^{lost}(n,o)$ and $\lambda(\delta)$ from the simulator of the environment given initial distributions of adversaries and obstacle density in each region (see Sec. \ref{sec:simulator} for more details).   

In this paper we aim to find a reactive control strategy for the vehicle.   A vehicle control strategy at a region $r$ depends on the facet $f$ through which the vehicle entered $r$.  It returns the facet $f'$ the vehicle should move towards, such that $(f,f')\in\Delta$.   The control strategy is reactive in the sense that it also depends on the number of adversaries and the obstacle density observed when entering the current region, as well as the movements of adversaries in the current region.  We are now ready to formulate the main problem we consider in this paper: 

\textbf{Problem:} Consider the partitioned environment defined by $R$ and $F$; initial facet and region $f_{init}$ and $r_{init}$; the motion capability $\Delta$ of a vehicle; initial adversary and obstacle density distributions for each region $p_r^{init}$ and $p_r^o$; the probability of losing the vehicle $p_{\delta}^{lost}$; rate of adversaries $\mu_l(r)$ and $\mu_e(r)$; and rate of the vehicle $\lambda(\delta)$; Find the vehicle control strategy that maximizes the probability of satisfying the Mission Objective.

The key idea of our approach is to model the motion of the vehicle in the environment, as well as vehicle estimates of adversary distributions in the environment as an MDP.   By capturing estimates of adversary distributions in this MDP, the vehicle updates the adversary distributions of its adjacent regions as it detects the movements of adversaries in the current region, and the control strategy produces an updated control if necessary.  As a result, a policy for the MDP is equivalent to a reactive control strategy for the vehicle in the environment.  We then translate the mission objective to a PCTL formula and find the optimal policy satisfying this formula with the maximum probability.

\begin{remark}
In this paper, we assume a ``deterministic" vehicle control model. In other words, we assume that the vehicle can use a motion primitive $(f,f')$ to move from facet $f$ to facet $f'$ of each region. We can easily extend the result of this paper to the case when the vehicle has a ``probabilistic" control model, in which the application of a motion primitive at a facet of a region enables transitions with known probabilities to several facets of the same region. This can easily achieved by modifying the transition probability function of the MDP.
\end{remark}
\section{Construction of an MDP Model}
\label{sec:MDP}
In this section we explain the construction of an MDP model for the motion of the vehicle and vehicle estimates of adversary distributions in the environment.  We first explain in Sec. \ref{sec:sub:update} how the vehicle updates its estimate of adversary distributions for its adjacent regions when an adversary enters or leaves the current region.  The updates of adversary distributions are captured in the MDP model.  Then we define the MDP in Sec. \ref{sec:sub:mdpconstruction}.  In Sec. \ref{sec:sub:P} we describe in detail how we obtain the transition probability function for the MDP model.  

\subsection{Update of the adversary distributions}
\label{sec:sub:update}
As adversaries enter and leave the current region, it is necessary to update the distributions of adversaries in adjacent regions.  Because the vehicle can only observe the movements of adversaries in its current region, and due to the assumption that distributions of adversaries in adjacent regions depend only on the current region and its adjacent regions, it is only necessary to update the adversary distributions for adjacent regions, and not for all regions in the environment.    Our MDP model captures all possible adversary distributions of adjacent regions at each region. 

Let us denote the distribution for region $r$ as $p_{r}$.    The initial adversary distribution of region $r$ is given in Eq. \ref{eq:probinit}.   Thus, the adversary distribution of region $r$ is a probability mass function $p_{r}:\{M,\ldots,N\}\rightarrow[0,1]$, where $M_{r}\leq M\leq N\leq N_{r}$. Note that if $M=N=N_r$, then $p_r(N_r)=1$ and no adversary may enter region $r$, or else the assumption that  $N_r$ is the maximum number of adversaries in region $r$ would be violated. Similarly, if $M=N=M_r$, then no adversary may leave region $r$.

Given the current adversary distribution $p_{r}$, assuming that an adversary has entered region $r$ (which means that $p_r(N_r)\neq1$), then we define the updated distribution as $p_{r}^{+}$ in the following way:
\begin{equation}
\label{eq:distri+range}
p_{r}^{+}:\left\{ \begin{array}{ll} \{M+1,\ldots,N\}\rightarrow[0,1] & \text{if } N=N_{r}  \\ \{M+1,\ldots,N+1\}\rightarrow[0,1] & \text{if } N<N_{r},  \end{array}\right.
\end{equation}
such that:
\begin{equation}
\label{eq:distri+value}
p_{r}^{+}(n)=\left\{ \begin{array}{ll}  p_{r}(n-1)+\frac{p_{r}(N)}{N-M} & \text{if } N=N_{r}  \\ p_{r}^{}(n-1) & \text{if } N<N_{r}.  \end{array}\right.
\end{equation}
Note that the probability distribution simply shifts by $1$ if $N<N_{r}$.  If $N=N_{r}$, given that an adversary entered region $r$, we can conclude that the previous number of adversaries cannot be $N_{r}$, thus we evenly redistribute the probability associated with $N_{r}$ before an adversary entered the region.

Similarly, assuming that an adversary has left region $r$, then $p_{r}(M_{r})\neq 1$ and we define the updated distribution as $p_{r}^{-}$ in the following way:
\begin{equation}
\label{eq:distri-range}
p_{r}^{-}:\left\{ \begin{array}{ll} \{M,\ldots,N-1\}\rightarrow[0,1] & \text{if } M=M_{r}  \\ \{M-1,\ldots,N-1\}\rightarrow[0,1] & \text{if } M>M_{r},  \end{array}\right.
\end{equation}
such that:
\begin{equation}
\label{eq:distri-value}
p_{r}^{-}(n)=\left\{ \begin{array}{ll}  p_{r}(n+1)+\frac{p_{r}(M)}{N-M} & \text{if } M=M_{r}  \\ p_{r}^{}(n+1) & \text{if } M>M_{r}.  \end{array}\right.
\end{equation}

\begin{figure}
            \centering
	\includegraphics[scale=0.31]{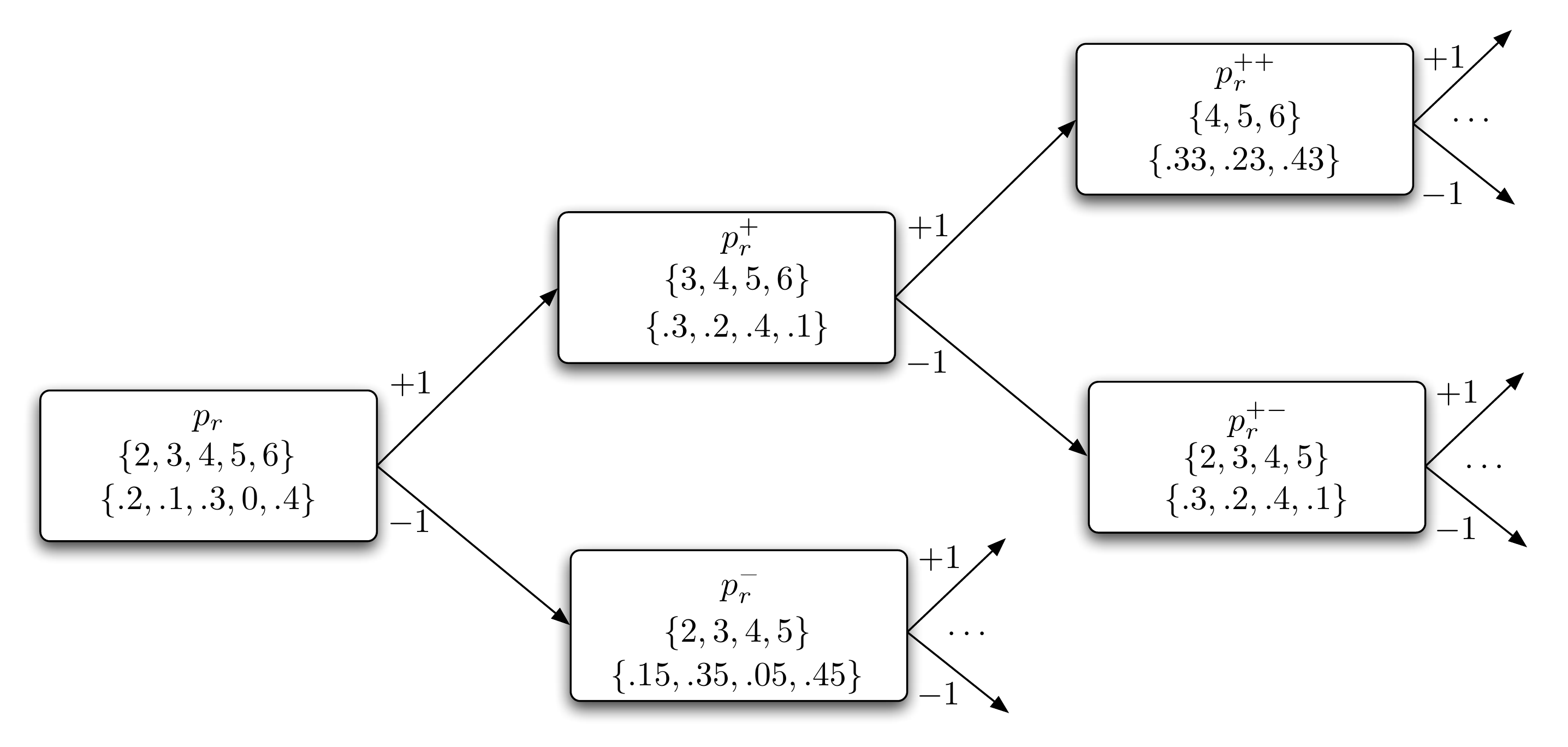}
      \caption{An example of using a tree to obtain all possible distributions for adversaries in a region.  Starting with $p_{r}^{init}$, we set $p_r=p_{r}^{init}$ and obtain all distributions in $D_{r}$ using Eq. (\ref{eq:distri+range})-(\ref{eq:distri-value}).  In this example, we have $p_{r}:\{2,3,4,5,6\}\rightarrow[0,1]$, where $p_{r}(2)=.2$, $p_{r}(3)=.1$, $p_{r}(4)=.3$, $p_{r}(5)=0$ and $p_r(6)=.4$.   Each box denotes the updated distribution from a previous distribution.  An arrow with $+1$ (or $-1$) means that we assume an adversary entered (or left) region $r$.  
      }  
            \label{fig:prob}
   \end{figure}

Given $p_{r}$, it is easy to verify that $p_{r}^{+}:\{M,\ldots,N\}\rightarrow[0,1]$ is a valid probability mass functions, {\it i.e.} $\sum_{n=M}^{N}p_{r}^{+}(n)=1$ (similarly for $p_{r}^{-}$).   Starting with the initial distribution $p_{r}^{init}$, we can use Eq. (\ref{eq:distri+range})-(\ref{eq:distri-value}) to determine all possible adversary distributions for region $r$.  We denote the set of all possible distributions for region $r$ as $D_{r}$.  We can use a tree to obtain $D_{r}$ with an example showing in Fig. \ref{fig:prob}.  

\subsection{MDP construction}
\label{sec:sub:mdpconstruction}
To begin the construction of the MDP model, we denote $B \subseteq F \times R$ as the boundary relation where $(f,r) \in B$ if and only if $f$ is a facet of region $r$.  
We denote the set of regions adjacent to region $r$ as $A_r=\{r_1,\ldots,r_{m}\} \subset R$.

Given $R$, $F$, $\Delta$, $D_{r}$, $p_r^o$, $p_{\delta}^{lost}$, $\mu_l(r)$, $\mu_e(r)$ and $\lambda(\delta)$, we define a labeled MDP $\mathcal{M}$ as a tuple $(S,s_0,Act, A,P,\Pi,h)$ (see Def. \ref{def:MDP}), where:
\begin{itemize}
\item $S=\bigcup_{r \in R}\{\{(f,z) \in B|z=r\} \times \{M_{r},\ldots,N_{r}\} \times  \{0,1,\ldots,N_r^o\} \times \text{\{lost,alive\}}\times \prod_{r'\in A_{r}}D_{r'}\}$. The meaning of the state is as follows: $((f,r),n,o,\text{alive},p_{r_{1}},\ldots,$\\$p_{r_{m}})$ means that the vehicle is at facet $f$, heading towards region $r$, and in region $r$ there are $n$ adversaries, $o$ obstacles, the vehicle is currently not lost, and the adversary distribution for the adjacent region $r_{i}\in A_{r}=\{r_{1},\ldots,r_{m}\}$ is $p_{r_{i}}$.  $((f,r),n,o,\text{lost}, p_{r_{1}},\ldots,p_{r_{m}})$ means that the vehicle did not make it to facet $f$ because it was lost in the previous region while heading towards $f$;
\item $s_0=((f_{init},r_{init}), 0, 0, \text{alive}, p^{init}_{r_1'},\ldots,p^{init}_{r_k'})$ is the initial state, where $A_{r_{init}}=\{r_{1}',\ldots,r_{k}'\}$;
\item $Act=\Delta \cup \tau$ is the set of actions, where $\tau$ is a dummy action when the vehicle is lost; 
\item $A$ is defined as follows: If the vehicle is alive, then $A(s)=\{(f,f')\in \Delta\}$, otherwise $A(s)=\tau$;
\item We describe how we generate the transition probability function $P$ in Sec. \ref{sec:sub:P};
\item $\Pi = \{r_{p}, r_{d}, \text{alive\}}$ is the set of properties;
\item $h$ is defined as follows: If $s=((f,r),n,o,b,p_{1},\ldots,p_{m})$, then $\{\text{alive}\}\in h(s)$ if and only if $b=\text{alive}$, $\{r_{p}\}\in h(s)$ if and only if $r=r_{p}$, and $\{r_{d}\}\in h(s)$ if and only if $r=r_{d}$. \end{itemize}

As the vehicle moves in the environment, it updates its corresponding state on $\mathcal M$.  The vehicle updates its state when:
\begin{itemize}
\item it reaches a facet $f$ and enters a region $r$, and observes the number of adversary $n$ and obstacle density $o$ in region $r$, then it updates its state to $((f,r),n,o,\text{alive},p_{r_{1}}^{init},\ldots, p_{r_{m}}^{init})$;
\item an adversary leaves the current region $r$ and moves into region $r'$, given the current adversary distribution of region $r'$ as $p_{r'}$, the vehicle updates this distribution to $p^{+}_{r'}$;
\item an adversary enters the current region $r$ from region $r'$, given the current adversary distribution of region $r'$ as $p_{r'}$, the vehicle updates this distribution to $p^{-}_{r'}$
\end{itemize}

Since actions of $\mathcal M$ consists of $\Delta$, $\mathcal M$ is designed so that its control policy can be directly translated to a reactive control strategy for the vehicle.   When the vehicle updates its state in $\mathcal M$, then the action $\delta \in \Delta$ at its current state determines the next facet the vehicle should move towards.

\subsection{Generating the transition probability function $P$}
\label{sec:sub:P}
In this subsection we describe in detail how we generate the transition probability function $P$ for the MDP model.  First, we define a random variable $e$ for the time in between a vehicle entering the current region $r$ at facet $f$, heading towards facet $f'$ and an event occurring, which can be: 1) an adversary entering the current region; 2) an adversary leaving the current region; or 3) the vehicle reaching facet $f'$.  

\begin{figure*}[ht]
            \centering
\includegraphics[scale=0.36]{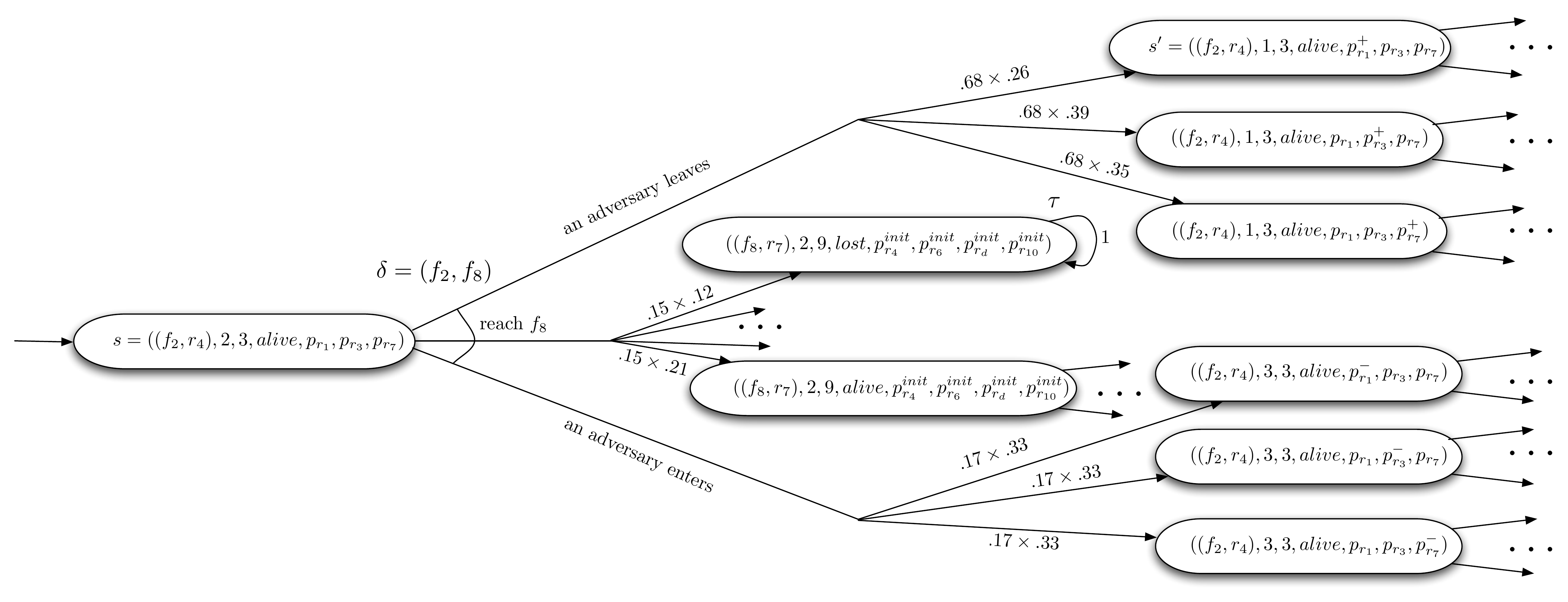}
      \caption{A fragment of the MDP $\mathcal{M}$ corresponding to the mission scenario shown in Fig. \ref{fig:partitioned_environment}.  As an example, assume the following: $\lambda((f_2,f_8))=.5$, $\mu_e(r_4)=\mu_l(r_4)=.3$,  $p_{r_1}(x)=p_{r_1}^{init}(x)=1/3,x\in\{1,2,3\}$, $p_{r_3}(x)=p_{r_3}^{init}(x)=1/3, x\in\{2,3,4\}$, $p_{r_7}(2)=p_{r_7}^{init}(2)=.6$, $p_{r_7}(3)=p_{r_7}^{init}(3)=.2$, $p_{r_7}(4)=p_{r_7}^{init}(4)=.2$.  The labels over the transitions correspond to the transition probabilities obtained using Eq. (\ref{eq:transition1})-(\ref{eq:transition3}) (\emph{e.g.}, the probability that an adversary leaves region $r_4$ is $.68$ and the probability that it enters region $r_1$ is $.26$, thus $P(s,\delta,s')=.68\times.26$).}
      \label{fig:MDPex} 
\end{figure*}

Note that if $X_1,\ldots,X_n$ are independent exponentially distributed random variables with rate parameters $\lambda_1,\ldots,\lambda_n$, then $min\{X_1,\ldots,X_n\}$ is exponentially distributed with parameter $\lambda=\sum_{i=1}^{n}\lambda_i$. The probability that $X_{k}$ is the minimum is $Pr(X_k=min\{X_1,\ldots,X_n\})=\frac{\lambda_k}{\lambda}$.  By assumption, movements of adversaries are independent of each other.  Since the arrival and departure of adversaries in the current region are modeled as two Poisson processes with inter-arrival and inter-departure time exponentially distributed with rate $\mu_{e}(r)$ and $\mu_{l}(r)$, respectively, and the time required for the vehicle to reach facet $f'$ is exponentially distributed with rate $\lambda(\delta)$, where $\delta=(f,f')$, the random variable $e$ is also exponentially distributed.  We assume $e$ is exponentially distributed with rate $\nu$.

At region $r$, assuming that the adversary distribution of an adjacent region $r'\in A_{r}$ is $p_{r'}$, we define $B_{r}\subseteq A_{r}$ as the set of adjacent regions $r'$ of $r$ such that $p_{r'}(M_{r'})\neq 1$ ({\it i.e.} the set of adjacent regions from which an adversary can leave) and $C_{r}\subseteq A_{r}$ as the set of adjacent regions $r'$ of $r$ such that $p_{r'}(N_{r'})\neq 1$ ({\it i.e.} the set of adjacent regions to which an adversary can enter).  We denote $E_{r}$ as the expected value for the distribution $p_{r}$.

Since the vehicle can not detect the exact number of adversaries in adjacent regions, only an estimated value $\nu_{e}$ of $\nu$ can be obtained from the expected number of adversaries in adjacent regions.  Assume the current state as $((f,r),n,o,\text{alive},p_{r_{1}},\ldots,p_{r_{m}})$.  If an adversary can leave current region $r$ ({\it i.e.} $n > M_r$ and $C_r \neq \emptyset$) then the time it takes for an adversary to leave region $r$ is exponentially distributed with rate $\mu_l(r)n$ because there are $n$ adversaries in the region and any of them can leave region $r$. Similarly, if an adversary can enter the current region $r$ ({\it i.e.} $n<N_r$), and there exists an adversary that can leave an adjacent region ({\it i.e.} $B_r \neq \emptyset$), then the time it takes for an adversary to enter region $r$ is exponentially distributed with the estimated rate $\mu_{e}(r)\sum_{r' \in B_r}E_{r'}$, where $\sum_{r' \in B_r}E_{r'}$ gives the total expected number of adversaries that can enter region $r$.  The time it takes for the vehicle to reach facet $f'$ is exponentially distributed with rate $\lambda(\delta)$.  Therefore, the estimated rate $\nu_{e}$ can be obtained as:  
\begin{equation}
\label{eq:nu}
\nu_{e}=\lambda(\delta)+\mu_l(r) n \mathbb I_{l}(A_{r},n)+\mu_e(r) \sum_{r'\in B_{r}}E_{r'} \mathbb I_{e}(n)
\end{equation}
where $n$ is the number of adversaries in the current region; $\mathbb I_{l}(A_{r},n)=0$ when $n=M_{r}$ or $C_{r}=\emptyset$, and $\mathbb I_{l}(A_{r},n)=1$ otherwise; and  $\mathbb I_{e}(n)=0$ if $n=N_{r}$, and $\mathbb I_{e}(n)=1$ otherwise. Indicator functions $\mathbb I_{l}(A_{r},n)$ and $\mathbb I_{e}(n)$ are used to determine if it is possible for an adversary to leave and enter the current region, respectively.  

The rate $\nu_{e}$ will be used to generate the probability transition function $P$.  We define the probability transition function $P: S \times Act \times S \rightarrow [0,1]$ as follows: 
\begin{itemize}
\item If $s=((f,r),n,o,\text{alive},p_{r_1},\ldots,p_{r_m})$, $s'=((f',r'),n',\\o',b',p^{init}_{r'_{1}},\ldots,p^{init}_{r'_k})$, with $\{r_{1},\ldots,r_{m}\}\in A_{r}$ and $\{r'_{1},\ldots,r'_{k}\}\in A_{r'}$, $\delta=(f,f')\in\Delta$ and $r'\in A_{r}$, then:
{\small
$P(s,\delta,s')=$
\begin{equation}\left\{
\begin{array}{ll}
\frac{\lambda(\delta)}{\nu_{e}}p_{r'}(n')p_{r'}^o(o')(1-p_{\delta}^{lost}(n,o)), & \text{ if } b'=\text{alive}\\ \frac{\lambda(\delta)}{\nu_{e}}p_{r'}(n')p_{r'}^o(o')p_{\delta}^{lost}(n,o), & \text{ if } b'=\text{lost}.
\end{array}\right.
\label{eq:transition1}
\end{equation}
}
Under the action $(f,f')$, the transition from state $s$ to $s'$ indicates that either the vehicle reaches facet $f'$ ($s'$ is an ``alive'' state) or the vehicle is lost while traversing the region $r$ ($s'$ is a ``lost'' state).   

Let us first consider the former case. $\frac{\lambda(\delta)}{\nu_{e}}$ corresponds to the probability that the vehicle reaches facet $f'$ before any adversary entering or leaving region $r$.  $p_{r'}(n')$ corresponds to the probability of observing $n'$ adversaries in region $r'$ when entering region $r'$ from facet $f'$.  $p_{r'}^o(o')$ corresponds to the probability of observing obstacle density $o'$ for region $r'$ when entering $r'$.  $(1-p_{\delta}^{lost}(n,o))$ corresponds to the probability of safely crossing the current region with $n$ adversaries and obstacle density $o$.   Since each of these events are independent with each other, the probability of transition is the multiplication of the above probabilities.  The same reasoning applies to the latter case, where $(1-p_{\delta}^{lost}(n,o))$ is replaced by $p_{\delta}^{lost}(n,o)$ as the probability of losing the vehicle while crossing region $r$.

\item If $s=((f,r),n,o,\text{alive},p_{r_1},\ldots,p_{r_m})$, $s'=((f,r),n+1,o,\text{alive},p_{r_1}',\ldots,p_{r_m}')$, with $\{r_{1},\ldots,r_{m}\}\in A_{r}$, $\delta=(f,f')\in\Delta$ for some $f'$, $p_{r_i}=p_{r_i}'$ for all $i=\{1,\ldots,m\}\setminus \{j\}$ and $p_{r_j}'=p_{r_j}^{-}$ for some $j$, then:  
\begin{equation}
\label{eq:transition2}
P(s,\delta,s')= \frac{\mu_e(r) E_{r_{j}}}{\nu_{e}}.
\end{equation}
The transition from state $s$ to $s'$ indicates that an adversary from region $r_{j}$ enters the current region.  Thus, the adversary distribution of region $r_{j}$ is updated to $p_{r_j}'=p_{r_j}^{-}$ (while the distributions for the other regions remain the same).  $\frac{\mu_e(r) E_{j}}{\nu_{e}}$ corresponds to the probability that an adversary enters region $r$ from $r_{j}$ before the vehicle reaches facet $f'$ or an adversary moves in between the current region and another adjacent region.  

\item If $s=((f,r),n,o,\text{alive},p_{r_1},\ldots,p_{r_m})$, $s'=((f,r),n-1,o,\text{alive},p_{r_1}',\ldots,p_{r_m}')$, with $\{r_{1},\ldots,r_{m}\}\in A_{r}$, $\delta=(f,f')\in\Delta$ for some $f'$,  $p_{r_i}=p_{r_i}'$ for all $i=\{1,\ldots,m\}\setminus \{j\}$ and $p_{r_j}'=p_{r_j}^{+}$ for some $j$, then:  
\begin{equation}
\label{eq:transition3}
P(s,\delta,s')=\frac{\mu_l(r) n}{\nu_{e}|C_{r}|},
\end{equation}
where $|C_{r}|$ is the cardinality of $C_{r}$.
The transition from the state $s$ to $s'$ indicates that an adversary leaves the current region and enters region $r_{j}$.  Thus, the adversary distribution of region $r_{j}$ is updated to $p_{r_j}'=p_{r_j}^{+}$.  $\frac{\mu_l(r) n}{\nu_{e}|C_{r}|}$ corresponds to the probability that an adversary enters $r_{j}$ from region $r$ before the vehicle reaches facet $f'$ or an adversary enters the current region.
\item If $s=((f,r),n,o,\text{lost},p_{r_1},\ldots,p_{r_m})$, then $P(s,\tau,s)=1$.  $s$ corresponds to the case where the vehicle is lost, thus it self-loops with probability $1$.
\item Otherwise, $P(s,\delta,s')=0$.
\end{itemize}

To help understand the computation of $P$, a fragment of the MDP model corresponding to the mission scenario in Fig. \ref{fig:partitioned_environment} is shown in Fig. \ref{fig:MDPex}.  The following proposition ensures that $P$ is a valid probability transition function:
\begin{prop}
$P$ is a valid probability transition function, {\it i.e.} $\sum_{s'\in S}P(s,\delta,s')=1$ if $\delta \in A(s)$ and $P(s,\delta,s')=0$ if $\delta\notin A(s)$.
\end{prop}
\begin{pf}
From the definitions of $P$ and $\mathcal{M}$ it follows that $P(s,\delta,s')=0$ if $\delta\notin A(s)$. We want to show that $\sum_{s'\in S}P(s,\delta,s')=1$ for all combinations of $B_r$, $C_r$, $M_r$, $N_r$, and $n$ when $\delta \in A(s)$.
Let us denote $S_t \subseteq S$ as the set of states that are defined as $s'$ in Eq. (\ref{eq:transition1}), $S_e \subseteq S$ as the set of states that are defined as $s'$ in Eq. (\ref{eq:transition2}) and $S_l \subseteq S$ as the set of states that are defined as $s'$ in Eq. (\ref{eq:transition3}).

If $B_r \neq \emptyset$ and $n<N_r$ with $C_r=\emptyset$ or $n=M_r$, then, by Eq. (\ref{eq:nu}), $\nu_e=\lambda(\delta)+\mu_e(r)\sum_{r' \in B_r}E_{r'}$. Using Eq. (\ref{eq:transition1})-(\ref{eq:transition3})  it follows that:
{\small
\begin{align*}
\sum_{s'\in S}P(s,\delta,s') &= \sum_{s' \in S_t}\frac{\lambda(\delta)}{\nu_{e}}p_{r'}(n')p_{r'}^o(o')p_{\delta}^{lost}(n,o)+\\ 
&+ \sum_{s' \in S_t}\frac{\lambda(\delta)}{\nu_{e}}p_{r'}(n')p_{r'}^o(o')(1-p_{\delta}^{lost}(n,o))+ \\
&+ \sum_{{s' \in S_e}} \frac{\mu_e(r) E_{r'}}{\nu_{e}}\\
&=\frac{\lambda(\delta)}{\nu_e}{\nu_e}\sum_{s' \in S_t}p_{r'}(n')p_{r'}^o(o')+\sum_{r' \in B_r} \frac{\mu_e(r) E_{r'}}{\nu_{e}}\\
&=\frac{\lambda(\delta)+\mu_e(r)\sum_{r' \in B_r}E_{r'}}{\nu_e}=1
\end{align*}}
Similarly, if $C_r \neq \emptyset$ and $n>M_r$ with $B_r=\emptyset$ or  $n=N_r$, then $\nu_e=\lambda(\delta)+\mu_l(r)n$, and using Eq. (\ref{eq:transition1})-(\ref{eq:transition3})  it follows that: 
{\small
\begin{align*}
\sum_{s'\in S}P(s,\delta,s')&= \sum_{s' \in S_t}\frac{\lambda(\delta)}{\nu_{e}}p_{r'}(n')p_{r'}^o(o')p_{\delta}^{lost}(n,o)+\\
\end{align*}
\begin{align*}
&+ \sum_{s' \in S_t}\frac{\lambda(\delta)}{\nu_{e}}p_{r'}(n')p_{r'}^o(o')(1-p_{\delta}^{lost}(n,o))+ \\
&+ \sum_{s' \in S_l}\frac{\mu_l(r) n}{\nu_{e}|C_{r}|} \\
&= \frac{\lambda(\delta)}{\nu_e}+|C_r|\frac{\mu_l(r) n}{\nu_{e}|C_{r}|} = \frac{\lambda(\delta)+\mu_l(r)n}{\nu_e}=1
\end{align*}
}
If $B_r=\emptyset$ and $C_r = \emptyset$ or if $n=M_r=N_r$, then $\nu_e=\lambda(\delta)$. Using Eq. (\ref{eq:transition1})-(\ref{eq:transition3})  it follows that:
{\small
\begin{align*}
\sum_{s'\in S}P(s,a,s')&= \sum_{s' \in S_t}\frac{\lambda(\delta)}{\nu_{e}}p_{r'}(n')p_{r'}^o(o')p_{\delta}^{lost}(n,o)+\\
&+ \sum_{s' \in S_t} \frac{\lambda(\delta)}{\nu_{e}}p_{r'}(n')p_{r'}^o(o')(1-p_{\delta}^{lost}(n,o)) \\
&=\frac{\lambda(\delta)}{\nu_e}=1.
\end{align*}
}
In the most general case, when $B_r \neq \emptyset$, $C_r \neq \emptyset$, and $M_r<n<N_r$, then, by Eq. (\ref{eq:nu}),  $\nu_e=\lambda(\delta)+\mu_l(r)n+\mu_e(r)\sum_{r' \in B_r}E_{r'}$. Using Eq. (\ref{eq:transition1})-(\ref{eq:transition3}) it follows that:
{\small
\begin{align*}
\sum_{s'\in S}P(s,\delta,s') &= \sum_{s' \in S_t}\frac{\lambda(\delta)}{\nu_{e}}p_{r'}(n')p_{r'}^o(o')p_{\delta}^{lost}(n,o) + \\
&+ \sum_{s' \in S_t}\frac{\lambda(\delta)}{\nu_{e}}p_{r'}(n')p_{r'}^o(o')(1-p_{\delta}^{lost}(n,o))+  \\
& + \sum_{s' \in S_l}\frac{\mu_l(r) n}{\nu_{e}|C_{r}|}+\sum_{{s' \in S_e}} \frac{\mu_e(r) E_{r'}}{\nu_{e}} \\
&= \frac{\lambda(\delta)+\mu_l(r)n+\mu_e(r)\sum_{r' \in B_r}E_{r'}}{\nu_e}=1
\end{align*}
}
Thus the proof is completed. \qed
\end{pf}

\section{Generating the optimal control policy and a vehicle control strategy}
\label{sec:vehicle_control}
After obtaining the MDP model, we solve our proposed problem by using the PCTL control synthesis approach presented in \citep{lahijanian:motion} by translating the problem to a PCTL formula.  The Mission Objective is equivalent to the temporal logic statement ``eventually reach $r_p$ and then $r_d$ while always staying alive'', which can be translated to the following formula $\phi$:
{\small
\begin{equation}
\label{eq:PCTLformula}
\mathcal{P}_{max=?}[ \text{alive} \; \mathcal{U} \;  (\text{alive} \wedge r_p \wedge \mathcal{P}_{>0}[\text{alive} \;  \mathcal{U} \; (\text{alive} \wedge r_d)])].
\end{equation}
}
Because formula $\phi$ has two temporal operators $\mathcal{U}$ (until), two maximum reachability probability problems (see \cite{baier:principles}) over the MDP need to be solved.   It should be noted that the nested $\mathcal{P}$-operator in formula $\phi$ (\emph{i.e.} $\mathcal{P}_{>0}[\Psi]$)  finds the control policy that maximizes the probability of satisfying $\Psi$ and returns all the initial states from which $\Psi$ is satisfied with probability greater than zero under this policy.

The PCTL control synthesis tool takes an MDP and a PCTL formula $\phi$ and returns the control policy that maximizes the probability of satisfying $\phi$ as well as the corresponding probability value by solving two linear programming problems. The tool is based on the off-the-shelf PCTL model-checking tool PRISM (see \cite{kwiatkowska:probabilistic}).   

We use Matlab to construct the MDP $\mathcal{M}$, which takes as input the partitioned environment defined by $R$ and $F$, the motion capability $\Delta$ of a vehicle and the values for $p^{init}_{r}$, $p_r^o$, $\mu_l(r)$ and $\mu_e(r)$ for all $r \in R$; and $p_{\delta}^{lost}$, $\lambda(\delta)$ for all $\delta \in \Delta$.  
Then the MDP $\mathcal{M}$ together with $\phi$ are passed to the PCTL control synthesis tool.
The output of the control synthesis tool is the optimal control policy that maximizes the probability of satisfying $\phi$.   This policy can be directly translated to the desired vehicle control strategy. 

The computational complexity of our approach is as follows: Given $R$, $A_r$, $N_r$, $M_r$, $N_r^o$ and $D_r$, the size of the MDP $\mathcal{M}$ is bounded above by $\max_{r \in R}(|B| \times (N_r-M_r+1) \times N_r^o \times |D_r| \times 2)$, where $|B|$ is bounded above by $|R| \times |A_r|$ and $|D_r|$ is bounded above by $((2(N_r-M_r)+1)^{|A_r|})$. The time complexity of the control synthesis tool is polynomial in the size of the MDP and linear in the number of the temporal operators. 

\section{Simulator of the Environment} 
\label{sec:simulator}
\begin{figure}[htp]
\begin{center}
  \includegraphics[scale=0.4]{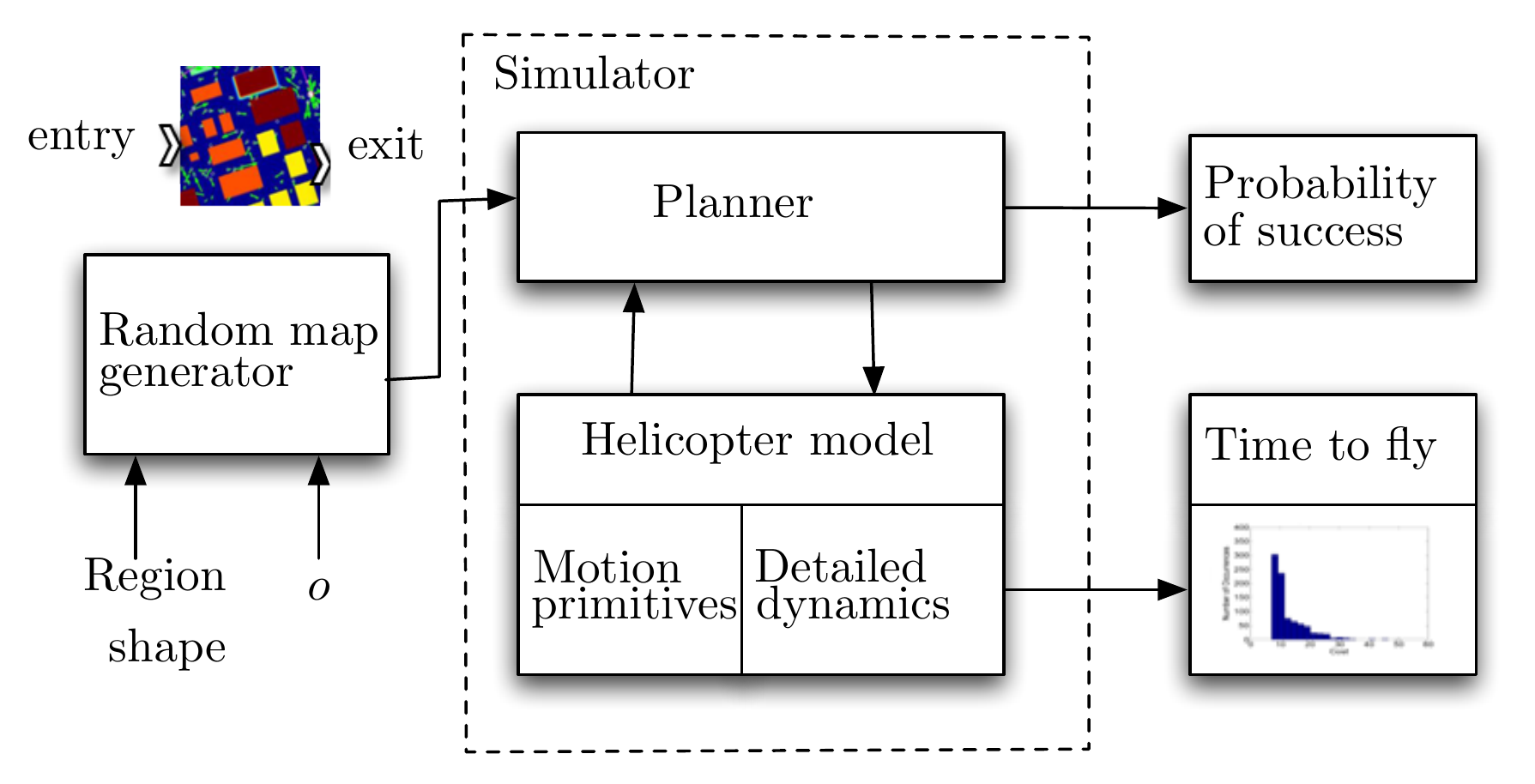}
  \caption{Test environment used to compute the probability
  $p_{\delta}^{lost}(n,o)$ and the rate $\lambda(\delta)$. }
  \label{fig:simulation_setup}
\end{center}
\end{figure}

We constructed a realistic test environment in order to obtain the probability $p^{lost}_{\delta}$ (Eq. \ref{eq:probkill}) from existing data of the distribution of obstacles in each region, and values for rate of the vehicle, $\lambda(\delta),\delta\in\Delta$.  This test environment consists of several components, which are shown in Fig.
\ref{fig:simulation_setup}.

In order to obtain $p^{lost}_{\delta}$, we first generated the marginal probability $p_{\delta}^{lost}(o)$, $\delta=(f,f')$ as the probability of losing the vehicle while traversing region $r$ from facet $f$ to $f'$ with obstacle density $o$. 
This probability depends on the motion planning algorithm for the vehicle traversing the region, and the ability of the vehicle to detect obstacles.  We assumed that the obstacle data in the environment was accurate and that there was no need for real-time obstacle detection. We used a probabilistic road-map planner \citep{lavalle:planning,frewen:adaptive} to solve the following problem: 
given a starting point on a facet $f$ and an ending point on the facet $f'$, find a shortest collision free path between them.
The planner uses a randomized algorithm that consists of building a random graph over the free space in the environment, and finding the shortest feasible collision-free path.  Because of the randomized nature of the algorithm, there is a non-zero probability that a path can not be found by the planner even if one exists. This is the probability $p_{\delta}^{lost}(o)$ because it is the probability that the vehicle can not safely traverse from facet $f$ to $f'$.

We computed $p_{\delta}^{lost}(o)$ using sampling (Fig.~\ref{fig:simulation_setup}). The random parameters that we considered were the size and position of objects in a region. Specifically, given the obstacle density $o$, we generated a random map by instantiating obstacles with random positions and sizes so that the density was $o$. The map was provided to the planner that generated a path. We used a symbolic control approach to plan the motion of the vehicle in the environment. Specifically, to implement the planner at the top of Fig. \ref{fig:simulation_setup}, we used the vehicle motion primitives defined in \citep{frazzoli:maneuver}. The successes and failures for each path were recorded. When a feasible path was found, a standard model of the dynamics of a helicopter \citep{bullo:geometric} was used to simulate a trajectory following the path and compute $\lambda(\delta)$. 

We computed the joint probability $p_{\delta}^{lost}(n,o)$ as a combination of the marginal probabilities $p_{\delta}^{lost}(n)$ and $p_{\delta}^{lost}(o)$. The main reason for this approach was that while an accurate model is available to compute the probability of failing to traverse a region due to obstacles, the effect of adversaries is difficult to model and it is part of our future work. For the purposes of the case study in Sec. \ref{sec:examples}, we assumed the probability of losing the vehicle due to adversaries to be $p_{\delta}^{lost}(n)=0.01(n)^2$ for $n \in [0,10]$. After the marginal probabilities were obtained, we constructed the joint probability $p_{\delta}^{lost}(n,o)$ using the following formula (see \citep{nelsen:introduction}): 
\[
p_{\delta}^{lost}(n,o) = e^{-\sqrt{-log(p_{\delta}^{lost}(n)) -
log(p_{\delta}^{lost}(o)) }}.
\]

\section{Results}
\label{sec:examples}
We considered the scenario shown on Fig.~\ref{fig:real_scenario} together with the partitioned environment  and the possible motion of the vehicle $\Delta$ shown on Fig.~\ref{fig:partitioned_environment}. The initial probability mass function for adversaries in region $r \in R$, $p_r^{init}$, and the probability mass function of the obstacle density in region $r \in R$, $p_r^o$, are given in Table \ref{table:data}. In addition, we assumed that there is no adversary or obstacle in region $r_p$ and $r_d$. The probability $p_{\delta}^{lost}(n,o)$ and the rates of the vehicle $\lambda(\delta)$ for all $\delta \in A$ were obtained from the simulator. We used the following numerical values: $\lambda((f,f'))=0.128$ when $f$ and $f'$ are facets of $r_1$ and $r_5$, $\lambda((f,f'))=0.125$ when $f$ and $f'$ are facets of $r_2$, $r_4$, $r_8$, $r_9$, $r_{10}$, and $r_{11}$,  and $\lambda((f,f'))=0.091$ when $f$ and $f'$ are facets of $r_3$, $r_6$, and $r_7$ with $\mu_e(r)=\mu_l(r)=0.05$ for all $r\in R$.

\begin{table}[h]
\caption{Obstacle density and adversary distribution}
\label{table:data}
\begin{center}
\scalebox{0.85}{

\begin{tabular}{| c || c || c || c |}
\hline 
Region & Obstacle & \multicolumn{2}{c|}{Adversary distribution}\\ 
\cline{3-4}& density & case A	&  case B\\
\hline
$r_1$ & $1\%$ & $p_{r_1}^{init}(0)=1$ & $p_{r_1}^{init}(0)=1$\\
\hline
$r_2$ & $3\%$ & $p_{r_2}^{init}(x)=1/3,x \in [7,9]$ & $p_{r_2}^{init}(x)=1/3,x \in [2,4]$\\
\hline
$r_3$ & $6\%$ & $p_{r_3}^{init}(x)=1/3,x \in [7,9]$ & $p_{r_3}^{init}(x)=1/3,x \in [2,4]$ \\
\hline
$r_4$ & $5\%$ & $p_{r_4}^{init}(x)=1/3,x \in [1,3]$ & $p_{r_4}^{init}(x)=1/3,x \in [2,4]$\\
\hline
$r_5$ & $1\%$ & $p_{r_5}^{init}(x)=1/3,x \in [7,9]$ & $p_{r_5}^{init}(x)=1/3,x \in [2,4]$\\
\hline
$r_6$ & $9\%$ & $p_{r_6}^{init}(x)=1/3,x \in [7,9]$ & $p_{r_6}^{init}(x)=1/3,x \in [2,4]$\\
\hline
$r_7$ & $9\%$ & $p_{r_7}^{init}(x)=1/3,x \in [1,3]$ & $p_{r_7}^{init}(x)=1/3,x \in [2,4]$\\
\hline
$r_8$ & $3\%$ & $p_{r_8}^{init}(x)=1/3,x \in [1,3]$ & $p_{r_8}^{init}(x)=1/3,x \in [2,4]$\\
\hline
$r_9$ & $4\%$ & $p_{r_9}^{init}(x)=1/3,x \in [1,3]$ & $p_{r_9}^{init}(x)=1/3,x \in [4,6]$\\
\hline
$r_{10}$ & $4\%$ & $p_{r_{10}}^{init}(x)=1/3,x \in [1,3]$ & $p_{r_{10}}^{init}(x)=1/3,x \in [4,6]$\\
\hline
$r_{11}$ & $3\%$ & $p_{r_{11}}^{init}(x)=1/3,x \in [7,9]$ & $p_{r_{11}}^{init}(x)=1/3,x \in [2,4]$\\
\hline
\end{tabular}
}
\end{center}
\end{table}

\begin{figure}
            \centering
	\includegraphics[scale=0.415]{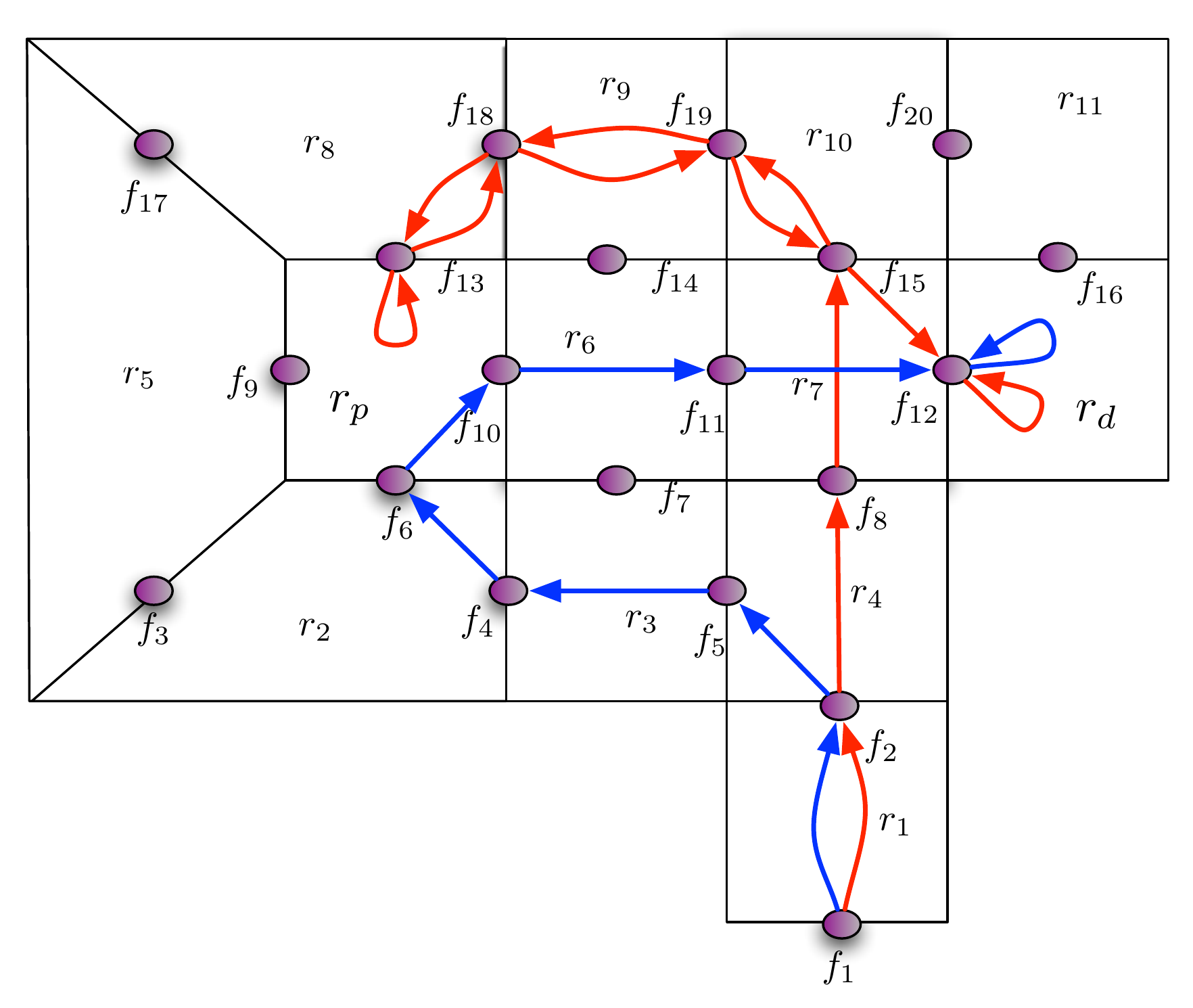}
      \caption{Runs of the vehicle in the partitioned environment for the given mission scenario and the data. Two different adversary distributions are given in Table \ref{table:data}.  The arrows represent movement of the vehicle in between facets. Red arrows correspond to case A, and  blue arrows correspond to case B.}
      \label{fig:Solution}
   \end{figure}

We obtained the vehicle control strategy through the method described in Sec. \ref{sec:vehicle_control}. 
Two vehicle runs are shown in Fig. \ref{fig:Solution}, corresponding to {case A} and {case B} (Table~\ref{table:data}).  We found that the maximum probability of satisfying the specification $\phi$ (Eq. \ref{eq:PCTLformula})  for {cases A} and {B} to be 0.141 and 0.805, respectively. The substantial difference between these two maximum probabilities is due to the difference in adversary distributions. A close analysis of the vehicle runs together with the adversary distributions shows that in {case A} the number of adversaries in regions $r_2$, $r_3$ and $r_6$ is high, which results in the vehicle control strategy that ensures that the vehicle avoids this regions. 

For this particular case study, the MDP $\mathcal{M}$ had 1079 states. The Matlab code used to construct $\mathcal{M}$ ran for approximately 14 minutes on a MacBook Pro computer with a 2.5 GHz dual core processor. Furthermore, the time it took the control synthesis tool to generate optimal policy is 4 minutes.

\section{Conclusions and Final Remarks}
\label{sec:conclusions}
In this paper we provided an approach to obtain a reactive control strategy that provides probabilistic guarantees for achieving a mission objective in a threat-rich environment.  We modeled the motion of the vehicle, as well as vehicle estimates of the adversary distributions as an MDP.   We then found the optimal control strategy for the vehicle maximizing the probability of satisfying a given mission task specified as a PCTL formula.  

Future work include extensions of this approach to a richer specification language such as probabilistic Linear Temporal Logic (PLTL) and a more general model of the vehicle in the environment such as a Partially Observed Markov Decision Process (POMDP).  
   
\bibliographystyle{alpha}        
\bibliography{References}          

\begin{thebibliography}{19}
\providecommand{\natexlab}[1]{#1}
\providecommand{\url}[1]{\texttt{#1}}
\expandafter\ifx\csname urlstyle\endcsname\relax
  \providecommand{\doi}[1]{doi: #1}\else
  \providecommand{\doi}{doi: \begingroup \urlstyle{rm}\Url}\fi

\bibitem[Alterovitz et~al.(2007)Alterovitz, Simeon, and
  Goldberg]{alterovitz:stochastic}
R.~Alterovitz, T.~Simeon, and K.~Goldberg.
\newblock The stochastic motion roadmap: A sampling framework for planning with
  markov motion uncertainty.
\newblock \emph{Proceedings of Robotics: Science and Systems}, 2007.

\bibitem[Baier et~al.(2008)Baier, Katoen, and Larsen]{baier:principles}
C.~Baier, J.~P. Katoen, and K.~M. Larsen.
\newblock \emph{Principles of Model Checking}.
\newblock MIT Press, 2008.

\bibitem[Bertsekas(1995)]{bertsekas:dynamic}
D.~Bertsekas.
\newblock \emph{Dynamic Programming and Optimal Control}.
\newblock Athena Scientific, 1995.

\bibitem[Bullo and Lewis(2004)]{bullo:geometric}
F.~Bullo and A.~D. Lewis.
\newblock \emph{Geometric Control of Mechanical Systems}, volume~49 of
  \emph{Texts in Applied Mathematics}.
\newblock Springer Verlag, New York-Heidelberg-Berlin, 2004.

\bibitem[Conner et~al.(2007)Conner, Kress-Gazit, Choset, Rizzi, and
  Pappas]{cooner:valet}
D.~C. Conner, H.~Kress-Gazit, H.~Choset, A.~Rizzi, and G.~J. Pappas.
\newblock Valet parking without a valet.
\newblock In \emph{Proceedings of 2007 IEEE/RSJ International Conference on
  Intelligent Robots and Systems}, pages 572--577, San Diego, CA, 2007.

\bibitem[Frazzoli et~al.(2005)Frazzoli, Dahleh, and Feron]{frazzoli:maneuver}
E.~Frazzoli, M.~A. Dahleh, and E.~Feron.
\newblock Maneuver-based motion planning for nonlinear systems with symmetries.
\newblock \emph{IEEE Trans. on Robotics}, 2005.

\bibitem[Frewen et~al.(2011)Frewen, Sane, Kobilarov, Bajekal, and
  Chevva]{frewen:adaptive}
T.~A. Frewen, H.~Sane, M.~Kobilarov, S.~Bajekal, and K.~R. Chevva.
\newblock Adaptive path planning in a dynamic environment using a receding
  horizon probabilistic roadmap.
\newblock In \emph{AHS International Specialists' Meeting}, 2011.

\bibitem[Karaman and Frazzoli(2008)]{karaman:vehicle}
S.~Karaman and E.~Frazzoli.
\newblock Vehicle routing problem with metric temporal logic specifications.
\newblock In \emph{IEEE Conf. on Decision and Control}, 2008.

\bibitem[Kloetzer and Belta(2008{\natexlab{a}})]{kloetzer:dealing}
M.~Kloetzer and C.~Belta.
\newblock Dealing with nondeterminism in symbolic control.
\newblock In \emph{Hybrid Systems: Computation and Control: 11th International
  Workshop}, 2008{\natexlab{a}}.

\bibitem[Kloetzer and Belta(2008{\natexlab{b}})]{kloetzer:fully}
M.~Kloetzer and C.~Belta.
\newblock A fully automated framework for control of linear systems from
  temporal logic specifications.
\newblock \emph{IEEE Transactions on Automatic Control}, 2008{\natexlab{b}}.

\bibitem[Kress-Gazit et~al.(2007)Kress-Gazit, Fainekos, and
  Pappas]{kress-gazit:whereswaldo?}
H.~Kress-Gazit, G.~E. Fainekos, and G.~J. Pappas.
\newblock Where's waldo? sensor-based temporal logic motion planning.
\newblock In \emph{In IEEE International Conference on Robotics and
  Automation}, pages 3116--3121, 2007.

\bibitem[Kwiatkowska et~al.(2004)Kwiatkowska, Norman, and
  Parker]{kwiatkowska:probabilistic}
M.~Kwiatkowska, G.~Norman, and D.~Parker.
\newblock Probabilistic symbolic model checking with {PRISM}: A hybrid
  approach.
\newblock \emph{International Journal on Software Tools for Technology
  Transfer}, 6\penalty0 (2):\penalty0 128--142, 2004.

\bibitem[Lahijanian et~al.(2010)Lahijanian, Wasniewski, Andersson, and
  Belta]{lahijanian:motion}
M.~Lahijanian, J.~Wasniewski, S.~B. Andersson, and C.~Belta.
\newblock Motion planning and control from temporal logic specifications with
  probabilistic satisfaction guarantees.
\newblock In \emph{IEEE International Conference on Robotics and Automation},
  pages 3227--3232, 2010.

\bibitem[LaValle(2006)]{lavalle:planning}
S.~M. LaValle.
\newblock \emph{Planning Algorithms}.
\newblock Cambridge University Press, Cambridge, U.K., 2006.

\bibitem[Loizou and Kyriakopoulos(2004)]{loizou:automatic}
S.~G. Loizou and K.~J. Kyriakopoulos.
\newblock Automatic synthesis of multi-agent motion tasks based on ltl
  specifications.
\newblock In \emph{43rd IEEE Conference on Decision and Control}, pages
  153--158, 2004.

\bibitem[Nelsen(2006)]{nelsen:introduction}
R.~Nelsen.
\newblock An introduction to copulas.
\newblock Springer-Verlag New York, Inc., 2006.

\bibitem[Ross(2006)]{ross:introduction}
S.~Ross.
\newblock \emph{Introduction to Probability Models}.
\newblock Academic Press, Inc., 2006.

\bibitem[Rutten et~al.(2004)Rutten, Kwiatkowska, Norman, and
  Parker]{rutten:mathematical}
J.~Rutten, M.~Kwiatkowska, G.~Norman, and D.~Parker.
\newblock \emph{Mathematical techniques for analyzing concurrent and
  probabilistic systems}.
\newblock American Mathematical Society, 2004.

\bibitem[Topcu et~al.(2009)Topcu, Wongpiromsarn, and Murray]{topcu:receding}
U.~Topcu, T.~Wongpiromsarn, and R.~M. Murray.
\newblock Receding horizon temporal logic planning for dynamical systems.
\newblock In \emph{Proceedings of the 48th IEEE Conference on Decision and
  Control}, 2009.

\end{thebibliography}
\end{document}